\documentclass[twocolumn,aps,prb]{revtex4}

\usepackage{epsf}
\usepackage{bm}

\begin{document}

\title{Ab initio Molecular Dynamical Investigation of the Finite Temperature Behavior of
the Tetrahedral Au$_{19}$ and Au$_{20}$ Clusters.}

\author{Sailaja Krishnamurty, Ghazal Shafai and D. G. Kanhere}

\affiliation{ Department of Physics and Centre for Modeling and Simulation, University
of Pune, Ganeshkhind, Pune--411 007, India}

\author{B. Soul\'{e} de Bas and M. J. Ford}

\affiliation{Institute for Nanoscale Technology, University of Technology, Sydney
PO Box 123, Broadway, NSW 2007 Australia}

\begin{abstract}

Density functional molecular dynamics simulations have been carried out to understand 
the finite temperature behavior of Au$_{19}$ and Au$_{20}$ clusters.  
Au$_{20}$ has been reported to be a unique molecule having tetrahedral geometry, a large HOMO-LUMO energy gap
and an atomic packing similar to that of the bulk gold (J. Li et al., Science, {\bf 299} 864, 2003). Our results show
that the geometry of Au$_{19}$ is exactly identical to that of Au$_{20}$ with one missing corner atom (called as vacancy).
Surprisingly, our calculated heat capacities for this nearly identical pair of gold cluster exhibit dramatic differences. 
Au$_{20}$ undergoes a clear and distinct solid like to liquid like
transition with a sharp peak in the heat capacity curve around 770~K. On the other hand, Au$_{19}$ has a 
broad and flat heat capacity curve with continuous melting transition.
This continuous melting transition turns out to be a consequence of a process involving series of atomic
rearrangements along the surface to fill in the missing corner atom. 
This results in a restricted diffusive motion of atoms 
along the surface of Au$_{19}$ between 650~K to 900~K during which the shape of the ground state geometry is retained. 
In contrast, the tetrahedral structure of Au$_{20}$ is destroyed around 800~K, and the cluster is clearly in
a liquid like state above 1000~K. Thus, this work clearly demonstrates that (i) the gold clusters
exhibit size sensitive variations in the heat capacity curves and   
(ii) the broad and continuous melting transition in a cluster, 
a feature which has so far been attributed to the disorder or absence of symmetry in
the system, can also be a consequence of a defect (absence of a cap atom) in the
structure.

\end{abstract}


\maketitle
\section{Introduction}
\label{sec.intro}

Recently discovered clusters and nanostructures of gold are found
to have a rich chemistry with potential applications in materials science,~\cite{ms} medicine,~\cite{md}
and in the area of catalysis.~\cite{homo,hetro}
In particular, small clusters of gold have attracted interest as tips
and contacts in molecular electronic circuits~\cite{tips} and also
as chemical catalysts.~\cite{cata} Experimentally, even a small cluster such as Au$_{8}$ 
has been reported to catalyze the oxidation reaction of
CO.~\cite{pekka} However, these properties are reported to have 
strong size sensitive variations. Another factor influencing the application of
gold clusters is their thermal stability. It is noted that
several Au clusters undergo structural transformation or tend
to grow readily by migrating and merging~\cite{merge} under high-temperature
conditions (500~K and above). These effects have important consequences
in the applications involving elevated temperatures and the growth
mechanisms of clusters. In this context, a study on the finite temperature behavior of
Au clusters is of considerable importance. 

Since the pioneering reports on the possible applications of gold clusters,
a large amount of experimental~\cite{au-expt} and theoretical works~\cite{au-gold-theory} have been devoted to
understand the structural and electronic properties of Au$_{n}$ (n $\le$ 50) clusters.
These reports have demonstrated that gold clusters 
have very different physical and chemical properties as compared to their 
bulk counterpart. A more recent exciting report has shown photo electron
spectroscopic evidence of hollow golden cages with an average diameter of 5.5~\AA~in the
in the size range of 16 to 18 atoms.~\cite{cages} These predictions were 
further supported by the theoretical calculations in the same report.
However, Au$_{20}$ is the most intriguing gold cluster reported so far.~\cite{science} 
Experimental studies~\cite{cages,science} report this cluster to have a pyramidal structure 
(tetrahedral symmetry) with each of the four faces representing the 
(111) surface of the Face Centered Cubic (FCC) gold. 
It is reported to have a large energy gap between the Highest Occupied Molecular Orbital (HOMO) 
and Lowest Occupied Molecular Orbital (LUMO). This energy gap is greater than that of C$_{60}$ suggesting
it to be highly stable and chemically inert.
On the other hand, its structure with high surface area and large fraction of corner sites with low atomic coordination is
expected to provide ideal surface sites to bind various molecules such as CO, O$_{2}$ and CO$_{2}$
for catalysis. The structure of Au$_{19}$ is also seen to be very similar to that of Au$_{20}$ with one missing corner atom.
In this context, it is interesting to have an understanding on the thermal stability of these two 
gold clusters having an atomic packing similar to that of bulk gold. Hence, in the present work, we study the 
finite temperature behavior of Au$_{19}$ and Au$_{20}$ using the first principles Molecular Dynamics (MD) simulations.         

While several experimental and theoretical studies have been devoted to understand
the ground state geometries and chemical reactivity of gold clusters,
there are very few reports on the finite temperature properties of gold clusters.~\cite{Au-melting,Au} 
The classical MD simulations by Landman and co-workers~\cite{Au-melting,Au}
on medium sized Au clusters (150-1500 atoms) indicated that the
clusters in this size range undergo a solid--to--solid structural transformation 
around 700~K, before eventually melting
around 780~K. This is at a much lower value as compared
to the bulk melting temperature of 1377~K. 
However, to the best of our knowledge, only one ab initio molecular dynamics study~\cite{mike-melt}  
attempting to understand the finite temperature behavior
of Au clusters has been reported so far. 

As the cluster size reduces, the electronic effects play a
more explicit role in controlling the structural and thermal properties of the clusters. 
This is amply demonstrated by several 
first principles molecular dynamics simulations,~\cite{Ga-prl,Our-PRBsn10,Our-Silicon,James-Tin,Eur-Phys.J,Na-PRB,Na-JCP,Kavita-PRL,Aguado}
which have successfully explained various experimental 
findings~\cite{Haberland-1997,Haberland-Nature,Haberland-PRL-2003,Haberland-PRL-2005,Jarold-Tin,Jarrold-Gallium1,Jarrold-Jacs,Jarrold-Al,Jarrold-Tinfrag}
on the finite temperature behavior
of sodium, tin, gallium and aluminum clusters. 
These experimental studies have brought out various interesting phenomena such as
higher than bulk melting temperatures in Ga and Sn clusters,~\cite{Jarold-Tin,Jarrold-Gallium1}
and strong size dependent variations in the melting temperatures of Ga and Al clusters.~\cite{Jarrold-Jacs,Jarrold-Al}
However, the most surprising experimental finding is the 
size sensitive behavior of the shape of the heat capacities where
addition of even one atom is seen to result in a dramatic change of shape, prompting some of the clusters to be called as
``Magic Melters".~\cite{Jarrold-Jacs} This means that while some clusters do undergo a conventional and clear melting transition,
 others undergo a near continuous transition making it very
difficult to identify any meaningful transition temperature.
In a recent communication,~\cite{Kavita-PRL} we have clearly demonstrated that 
a cluster with local ``order" (an island
of atoms connected with equal bond strengths) displays a well
characterized melting transition with a distinct peak in the heat capacity
curve, while a {}``disordered\char`\"{} cluster is seen to undergo a continuous
transition with a flat heat capacity curve. 
Further, it is noted that this size sensitive nature
in small clusters is related to the evolutionary pattern
seen in their ground states and is seen to exist in clusters of sodium, gallium and aluminum.~\cite{Na-JCP,Ga-PRB,Al-pre}

In what follows, we show that this dramatic variation in the shape of heat capacity curve is
also observed in the present pair of gold clusters viz., Au$_{19}$ and Au$_{20}$. This observation has also
thrown light on additional factors responsible for a continuous melting transition in clusters.
As we shall see, in contrast to gallium or aluminum clusters, the flat or broad heat capacity curve in Au$_{19}$ is attributed
to a "vacancy" in the surface (or a surface defect). This ``vacancy" results in a chain of 
atomic rearrangements leading to a restricted diffusion of atoms along the surface. In contrast,
Au$_{20}$ undergoes a relatively sharp melting transition with a clear peak in the heat capacity curve around 770~K.
Thus, these results not only bring out an understanding on additional factors contributing to the broad melting transition in
clusters but also on the relative thermal stability of these unique tetrahedral gold clusters.

\section{Computational Details}
\label{sec.comp}

We have optimized about 300 geometries for each of the cluster, to obtain the ground state 
geometry and several low energy isomers. 
The initial configurations for the optimization
were obtained by carrying out a constant temperature dynamics of 100~ps
each at various temperatures between 400 to 1600~K. 
Once the ground state geometry is obtained, thermodynamic simulations are performed
using Born--Oppenheimer MD based
on the Kohn-Sham formulation of Density Functional Theory (DFT).\cite{KS} The ionic phase space
of the clusters is sampled classically in a canonical ensemble according
to the method proposed by N\'{o}se.~\cite{Nose}
The MD simulations have been carried
out using Vanderbilt's ultra soft pseudo potentials~\cite{uspp-vanderbilt}
within the Local Density Approximation (LDA) for describing the core-valance
interactions as implemented in the \textsc{vasp} package.~\cite{vasp}
Energy cutoff of 13.21~Ry is used for the plane wave expansion of
Au. We have used cubic super cells of length 20~\AA~and have ensured
that the results converge with respect to further increase in the
energy cutoff and size of the simulation box. 

In order to have a reliable sampling, we split the total temperature
range from 400--1600~K into at least 15 different temperatures for
both cluster sizes. We maintain the cluster at each temperature for
a period of at least 70~ps after equilibration, leading to a total simulation time of
around 1~ns. Following the finite-temperature study, the ionic heat
capacity of each cluster is computed using the multiple-histogram
(MH) method.\cite{MH,amv-review}
Various other thermodynamic indicators such as the mean-square 
displacements (MSD)'s of ions and the root-mean-square bond-length fluctuation
(RMS-BLF or $\delta_{{\rm rms}}$) are also computed. For the sake
of completeness we briefly discuss these parameters. The parameter
$\delta_{{\rm rms}}$ is a measure of the fluctuations in the bond
lengths averaged over all the atoms and over the total time span.
It is defined as \begin{equation}
\delta_{{\rm rms}}=\frac{2}{N(N-1)}\sum_{i>j}\frac{(\langle r_{ij}^{2}\rangle_{t}-\langle r_{ij}\rangle_{t}^{2})^{1/2}}{\langle r_{ij}\rangle_{t}},\label{eqn:delta}\end{equation}
 where $N$ is the number of atoms in the system, $r_{ij}$ is the
distance between atoms $i$ and $j$, and $\langle\ldots\rangle_{t}$
denotes a time average over the entire trajectory. The MSD is another
widely used parameter for analyzing a solid-like-to-liquid-like transition.
In the present work, we calculate the mean square displacement's
for individual atoms which is defined as \begin{equation}
\langle{\bf r}_{I}^{2}(t)\rangle=\frac{1}{M}\sum_{m=1}^{M}\left[{\bf R}_{I}(t_{0m}+t)-{\bf R}_{I}(t_{0m})\right]^{2},\label{eqn:msq}\end{equation}
where ${\bf R}_{I}$ is the position of the $I$th atom and we average
over $M$ different time origins $t_{0m}$ spanning the entire trajectory.
The MSD indicates the displacement of atom in the cluster as a function
of time. In the solid-like region, all atoms perform oscillatory motion
about fixed points resulting in a negligible MSD's of individual atoms
from their equilibrium positions. In a liquid-like
state, on the other hand, atoms diffuse throughout the cluster and
the MSD's eventually reach a saturated value of the order of the square
of the cluster radius. More technical details concerning the extraction
of thermodynamic averages, indicators and computation of the heat
capacity curve can be found in previous work.\cite{amv-review}

\section{Results and Discussion}
\label{sec.rd}

\begin{figure}
\epsfxsize=0.45\textwidth \centerline{\epsfbox{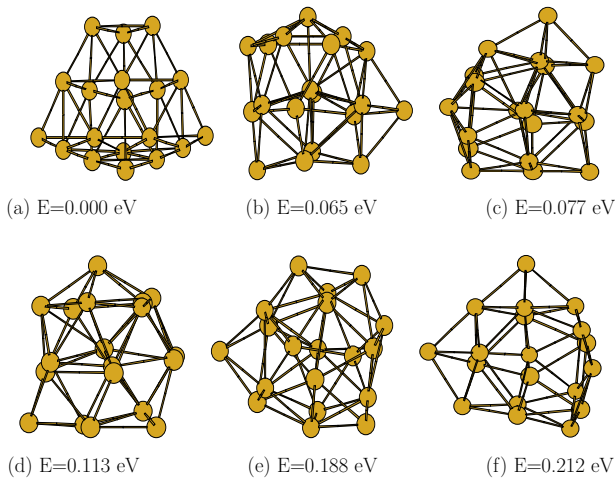}}
\caption{\label{Fig1} The ground state geometry and low lying isomers of
Au$_{19}$. The energy below is the difference in total energy of
the isomer with respect to the ground state energy.}
\end{figure}

\begin{figure}
\epsfxsize=0.45\textwidth \centerline{\epsfbox{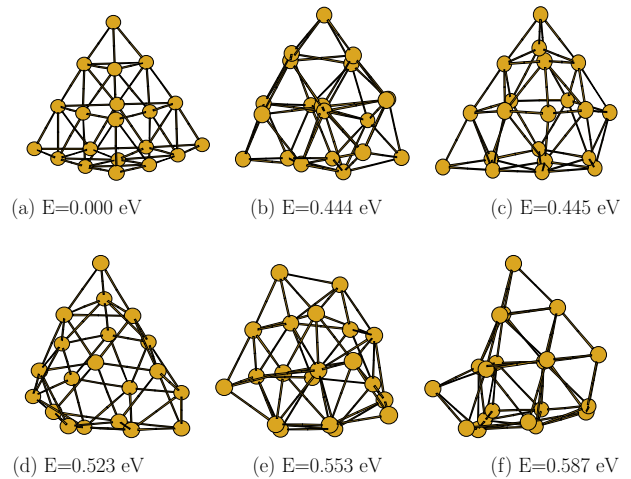}}
\caption{\label{Fig2} The ground state geometry and low lying isomers of
Au$_{20}$. The energy below is the difference in total energy of
the isomer with respect to the ground state energy.}
\end{figure}

We begin with a discussion on the ground state geometries
and some representative low lying isomers of Au$_{19}$ and Au$_{20}$ which are shown in the Fig.\ \ref{Fig1}
and Fig.\ \ref{Fig2}, respectively. It is clearly seen from these
figures that the ground state geometry of Au$_{20}$ (Fig.\ \ref{Fig2}--(a)) is a
tetrahedron. The ground state geometry of Au$_{19}$ (Fig.\ \ref{Fig1}--(a)) differs from that of Au$_{20}$ by a single
missing vertex atom of the tetrahedron. This is in agreement with
the recent experimental and theoretical predictions.~\cite{cages} Quite clearly both the structures
are symmetric, with ordered triangles stacked over one another.
The rest of the geometric parameters such as bond lengths, bond angles
and dihedral angles are almost identical in both the ground state geometries.
Thus, Au$_{19}$ can be considered as Au$_{20}$ with a vertex defect.

\begin{figure}
\epsfxsize=0.40\textwidth \centerline{\epsfbox{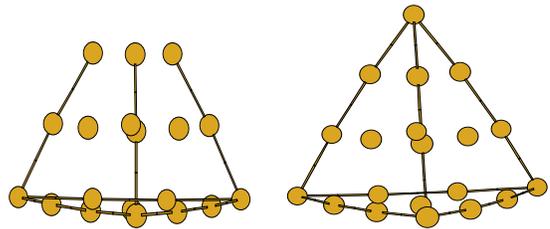}}
\caption{\label{Fig3} The distribution of shortest bonds (2.63~\AA) in
Au$_{19}$ and Au$_{20}$. Rest of the inter atomic bond distances (for the first nearest nearest neighbor) in both the
clusters are 2.75~\AA.}
\end{figure}

The atoms in Au$_{19}$ as well as Au$_{20}$ are bonded to their first nearest
neighbors with bond distances of either 2.63~\AA~(shortest bonds in the cluster) or 2.75~\AA~(next shortest
bonds). It is interesting to display
the connectivity of the shortest bonds to bring out the differences in them.
Fig.\ \ref{Fig3}--(a) and Fig.\ \ref{Fig3}--(b) show the distribution of shortest bonds
(2.63 \AA) for the case of Au$_{19}$ and Au$_{20}$, respectively. Clearly,
the shortest bonds are distributed only along the surface of both the clusters 
and form a closed network in Au$_{20}$, while in Au$_{19}$
they form an open network due to the missing atom.
It turns out, that the presence of the vertex defect
and open skeleton of shortest bonds in Au$_{19}$ play a significant role
in the finite temperature behavior of the cluster around 650~K and
initiate a set of restricted atomic rearrangements on the surface. 

The low lying isomers of Au$_{19}$ (shown in Fig.\ \ref{Fig1}--(b) to Fig.\ \ref{Fig1}--(f))
are clearly devoid of a regular
triangular arrangement of atoms seen in the ground state
configuration. The first low lying isomer of Au$_{19}$ (Fig.\ \ref{Fig1}--(b)) is nearly 0.065 eV higher
in energy as compared to the ground state configuration. Au$_{19}$ has several isomers with continuous energy
distribution between 0.065 eV to 0.7 eV some of which are shown in Fig.\ \ref{Fig1}--(c) to Fig.\ \ref{Fig1}--(f).
This can be contrasted with first low lying isomer of Au$_{20}$ (shown in Fig.\ \ref{Fig2}--(b)) 
which is almost 0.44 eV higher than the ground state
geometry. This structure has one central atom and rest of the 19 atoms arrange around this central atom so as
to have a highly deformed tetrahedron. This configuration is degenerate with the hollow cage configuration of
Au$_{20}$ (Fig.\ \ref{Fig2}--(c)).  
The fact that there are a couple of isomers well separated from the ground state geometry correlates very
well with the existence of a relatively sharp and well defined peak in the
heat capacity curve of Au$_{20}$.
Some other representative low lying isomers of Au$_{20}$ are
shown in Fig.\ \ref{Fig2}--(d) to Fig.\ \ref{Fig2}--(f). 
We also note that the atoms in all the high energy configurations of both clusters are bonded
to each other through a much wider and continuous range of Au-Au bond
lengths (ranging between 2.67~\AA -2.90~\AA) as compared to those
in the ground state geometry. 

\begin{figure}
\epsfxsize=0.5\textwidth \centerline{\epsfbox{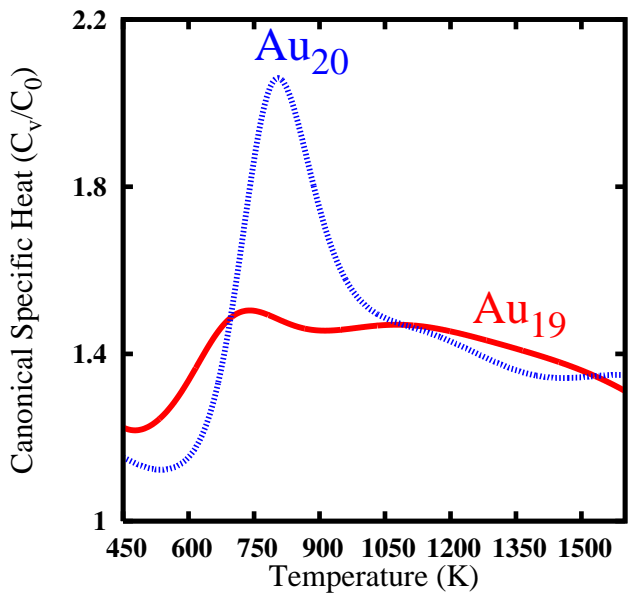}}
\caption{\label{Fig4} The Heat-Capacity Curves of Au$_{19}$ and Au$_{20}$.}
\end{figure}

\begin{figure}
\epsfxsize=0.5\textwidth \centerline{\epsfbox{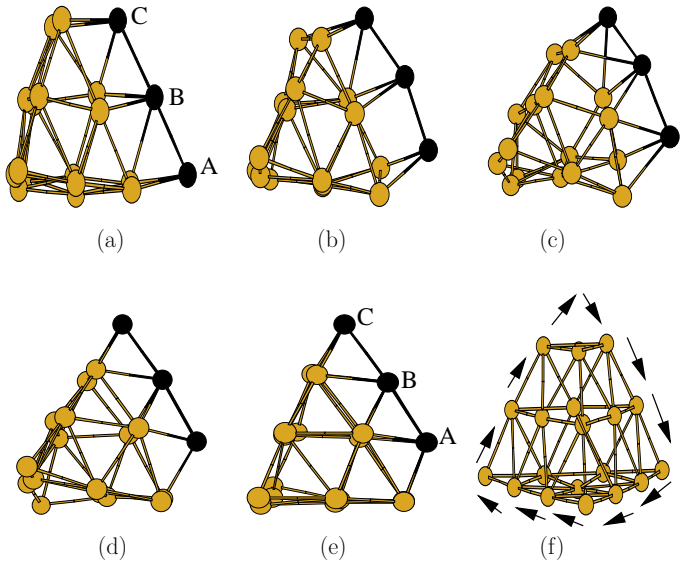}}
\caption{\label{Fig5} (a--e) Snapshots of restricted rearrangement of atoms in Au$_{19}$ around 650~K. (f)
Arrows depict the continuous atomic rearrangements that take place to fill in the missing cap atom between 700~K--900~K.}
\end{figure}

Now, we present the finite temperature behavior of both the clusters. 
we begin with a discussion on the calculated heat capacity curves
which is shown in Fig.\ \ref{Fig4}.
The figure brings out a remarkable feature, viz., a significant size sensitivity nature of
the heat capacity curves. The heat capacity curve of Au$_{20}$ has a 
clear and recognizable peak around 770~K, with a width of about 250~K.
In contrast, the heat-capacity curve of Au$_{19}$, a cluster with a vacancy, shows
a broad and almost continuous solid--to--liquid transition between
650 K--1200~K. Thus, this is yet an another example of dramatic change in the shape of
the heat capacity curve with the addition of a single atom. As already noted this
size sensitive nature has been observed earlier in Ga, Al clusters experimentally.~\cite{Jarrold-Jacs,Jarrold-Al}

It is possible to make a detailed analysis of the ionic motion by examining the trajectories of
the clusters. An analysis of the ionic motions of Au$_{19}$ reveals the cluster to vibrate
around its ground state geometry until 600~K. Around 650~K, the
cluster undergoes a peculiar structural rearrangement so as to fill in the
vacancy (the apex atom). The snapshots of this structural rearrangement
are shown from Fig.\ \ref{Fig5}--(a) to Fig.\ \ref{Fig5}--(e). 
Fig.\ \ref{Fig5}--(a) shows Au$_{19}$ with a missing cap atom. Note that,
at the end of the structural rearrangement (Fig.\ \ref{Fig5}--(e)), the vacancy which is present on the top in
Fig.\ \ref{Fig5}--(a) is shifted to the bottom edge. We denote the edge consisting of 
atoms `A', `B' and `C' as the reference edge. 
Coming to the details of the structural rearrangement, 
as the cluster evolves around 650~K, it is seen that these atoms in the reference 
edge push themselves upward (see Fig.\ \ref{Fig5}--(b) to Fig.\ \ref{Fig5}--(d)). The rest of atoms in the cluster
undergo minor displacements around their equilibrium positions during this process. At the end of this rearrangement
(see Fig.\ \ref{Fig5}--(e)), it is seen that 
the top edge atom (atom `C') in Au$_{19}$ moves to cap the missing vertex atom seen in Fig.\ \ref{Fig5}--(a).
The next edge atom (`B') moves up to
occupy the position initially occupied by atom `C' and atom the `A' occupies the 
position occupied by `B'. Now this creates a vacancy or defect at the position
initially occupied by the atom `A'. Thus, we now have a Au$_{19}$ cluster which is rotated by
90 degrees in the anti clock wise direction (Fig.\ \ref{Fig5}--(e)) with respect to Fig.\ \ref{Fig5}--(a). 

Around 650~K, only a single edge is displaced so as to cap the missing vertex atom.
Between 700~K--900~K, we see a continuous displacement of atoms along all the edges 
as shown in Fig.\ \ref{Fig5}--(f) as the vacancy is shifted from one vertex to the
other vertex.  A remarkable feature of this motion is that the overall shape of
the cluster remains approximately tetrahedron with a missing cap. Around 1000~K, the
tetrahedron structure is destroyed and the cluster visits its first and
second high energy configurations. The
cluster finally melts completely above 1200~K. This leads to a broad
feature (between 650~K--1200~K) in its heat capacity curve.

In contrast, the ionic motion of Au$_{20}$ shows all the atoms to
vibrate around their initial positions until about 750~K. 
The cluster undergoes a structural transformation from the ground state geometry
to the first isomer shown in Fig.\ \ref{Fig2}--(b) around 800~K.
The cluster visits other isomers around 900~K and melts completely above 1000~K leading to 
a clear and relatively narrow melting transition. 

\begin{figure}
\epsfxsize=0.5\textwidth \centerline{\epsfbox{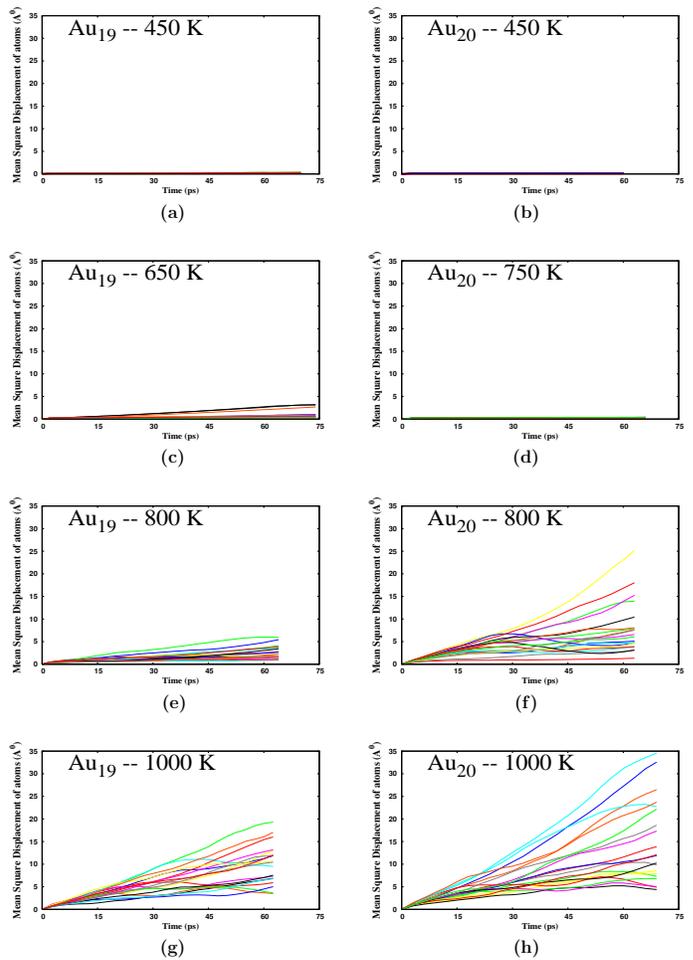}}
\caption{\label{Fig6} The Root Mean Square Displacements of atoms with respect
to the simulation time (ps) in Au$_{19}$ at various temperatures.}
\end{figure}

This contrasting behavior is brought out more clearly by examining the MSD's 
of the individual atoms. In Fig.\ \ref{Fig6}, we show the MSD's of individual atoms 
in both clusters in the temperature range of 450~K--1000~K. 
It is clearly from Fig.\ \ref{Fig6}--(a)
and Fig.\ \ref{Fig6}--(b) that atoms in both the clusters vibrate around their equilibrium 
positions at 450~K. Around 650~K, the rearrangement of atoms
along one edge in Au$_{19}$ is reflected in slightly higher 
mean square displacements (around 3~\AA) in Fig.\ \ref{Fig6}--(c). 
These values increase continuously in Au$_{19}$ as shown by the typical behavior around 800~K.
In contrast, the MSD values in Au$_{20}$ are negligible until 750~K ($<$ 0.5~$\AA$) and the values
increase sharply around 800~K. It is precisely at this temperature that the tetrahedron gets destroyed. 
The MSD's of Au$_{19}$ and Au$_{20}$ saturate around 35~\AA at 1200~K and 1000~K respectively 
indicating the presence of a liquid like state. 

\begin{figure}
\epsfxsize=0.5\textwidth \centerline{\epsfbox{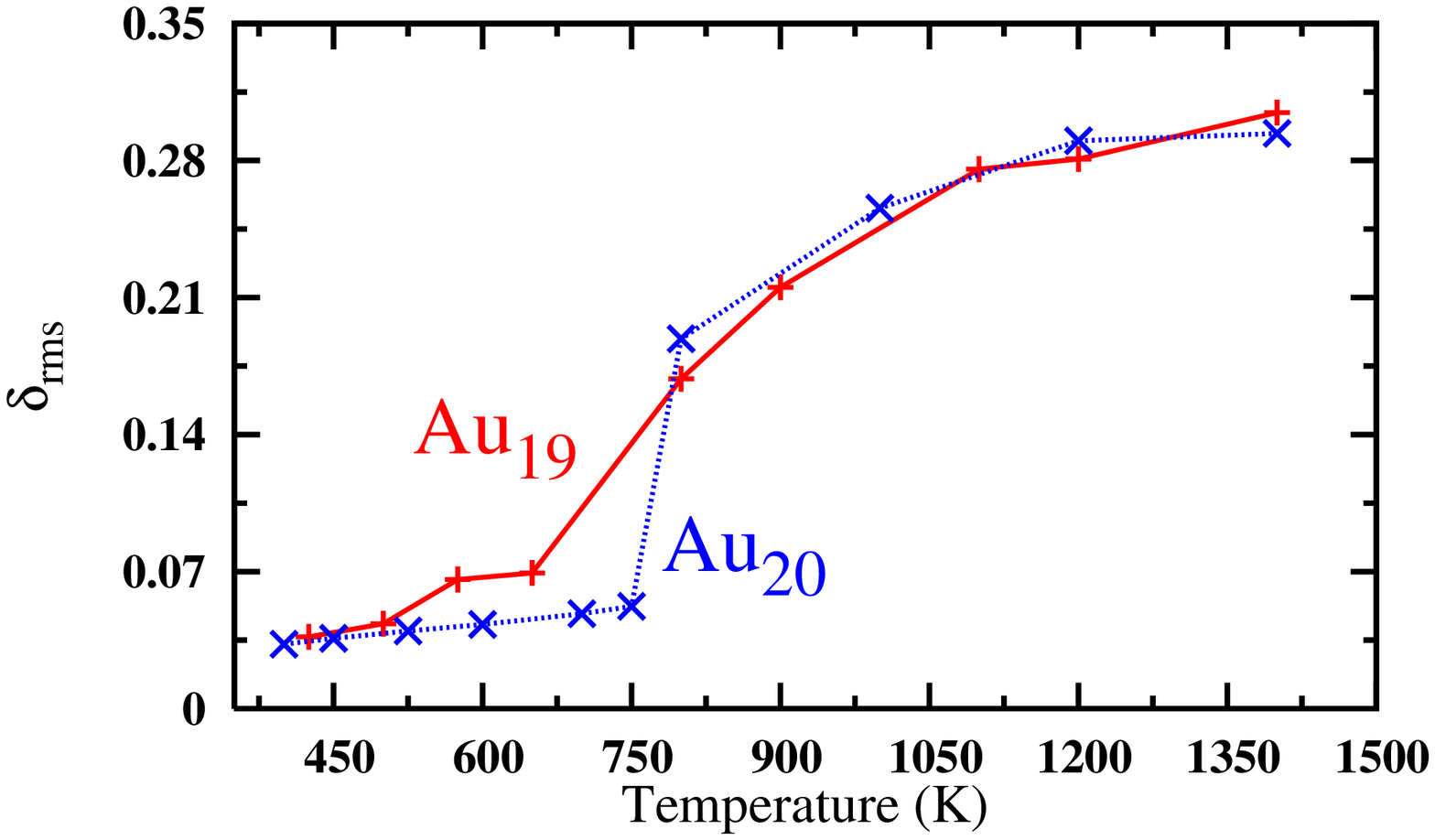}}
\caption{\label{Fig7} $\delta$$_{rms}$ of Au$_{19}$ and Au$_{20}$.}
\end{figure}

This contrasting behavior is some what weakly reflected in 
in average root-mean-square
bond-length fluctuation ($\delta_{{\rm rms}}$) of Au$_{19}$
and Au$_{20}$, which is in any case a quantity averaged out over all the atoms.
In Fig.\ \ref{Fig7} we show the $\delta_{{\rm rms}}$ for both the clusters.
As expected Au$_{20}$ shows a sharp transition indicated by a jump in 
$\delta_{{\rm rms}}$ value from 0.07 to 0.20.
\section{Summary and Conclusions}
\label{sec.intro}

In this work, we have presented the results of first principle molecular
dynamics simulations on Au$_{19}$ and Au$_{20}$, clusters which have atomic packing similar to that of the 
bulk gold. In spite of the fact, that the geometry 
of both the clusters is nearly identical except for a single vertex atom, 
they exhibit dramatic differences in the shape of their heat capacity curves.
We have shown that these differences are induced by vacancy. 
The vacancy in Au$_{19}$ induces a restricted diffusive motion along the surface of the 
cluster leading to a continuous melting transition.
In contrast, Au$_{20}$ exhibits a sudden and clear melting transition.
It may be noted that such a size sensitive nature of the heat capacity curves has been observed experimentally
in Al and Ga clusters and in Na clusters during the ab initio molecular dynamic simulations. 
In these studies the size sensitive nature of the heat capacity curves 
was attributed to the nature of the ``disorder" in the ground state
geometry. The present work clearly shows that this size sensitive behavior 
is also driven by the vacancy in the other wise perfect and symmetric ``lattice".
The work also demonstrates that the size sensitive 
variations in the melting characteristics to be generic in nature.
Finally, the contrasting finite temperature behavior reported in the present gold clusters
could have several implications in the applications of these clusters
and is a topic of further research interest.

\textbf{Acknowledgments}

GS and DGK thanks Indo French Center For Promotion of Advanced Research (IFCPAR--CEFIPRA)
for partial financial support (Project No. 3104-2). SK, GS and DSK thank Kavita Joshi for
useful discussions. MJF thanks the Australian Research Council (ARC) for financial support, the Australian
Centre for Advanced Computing and Communications(AC3) and Australian Partnership for
Advanced Computing (APAC) for computing facilities. BS thanks the
ARC for an International Postgraduate Research Scholarship (IPRS).


\begin{thebibliography}{20}
\bibitem{ms} Dyson, P. J.; Mingos, D. M. P. \textit{Gold. Progress
in Chemistry, Biochemistry and Technology} (ed. Schmidbaur, H.), Wiley,
Newyork, \textbf{1999}, 511.

\bibitem{md} Shaw III, C. F. \textit{Chem. Rev.} \textbf{1999}, 99,
2589.

\bibitem{homo} Teles, J. H.; Brode, S.; Chabanas, M. \textit{Angew.
Chem.} \textbf{1998}, 99, 2589.

\bibitem{hetro} Hashmi, A. S. K. \textit{Gold Bull.} \textbf{2003},
36, 3.

\bibitem{tips} Fan, F.-R. F.; Bard, A. J. \textit{Science} \textbf{1997},
277, 1791.

\bibitem{cata} Valden, M.; Lai, X.; Landman, D. W. \textit{Science}
\textbf{1998}, 281, 1647.

\bibitem{pekka} Pyykko, P. \textit{Angew. Chem. Int. Ed.} \textbf{2004},
43, 4412.

\bibitem{merge} Sutter, E.; Sutter, P.; Zhu, Y. \textit{Nano Letters}
\textbf{2005}, 5, 2092.

\bibitem{au-expt} Gilb, S.; Weis, P.; Furche, F.; Ahlrichs,
R.; Kappes, M. M. \textit{J. Chem. Phys.} \textbf{2002} 116, 4094;

\bibitem{au-gold-theory} Gilb, S.; Weis, P.; Furche, F.; Ahlrichs,
R.; Kappes, M. M. \textit{J. Chem. Phys.} \textbf{2002} 116, 4094;
Walker, A. V. \textit{J. Chem. Phys.} \textbf{2005} 122, 094310; Zhao,
J.; Yang, J.; Hou, J. G. \textit{Phys. Rev. B} \textbf{2003} 67, 085404;
Olson. R. M.; Varganov, S.; Gordon, M. S.; Metiu, H.; Chretien, S.;
Piecuch, P.; Kowalski, K.; Kucharski, S. A.; Musial, M. \textit{J.
Am. Chem. Soc.} \textbf{2005} 127, 1049; de Bas, B. S.; Ford, M. J.;
Cortie, M. B. \textit{J. Mol. Struc.} \textbf{2004} 686, 193; Mills,
G.; Gordon, M. S.; Metiu, H. \textit{J. Chem. Phys.} \textbf{2003}
118, 4198; Häkkinen, H.; Landman, U. \textit{Phs. Rev. B} \textbf{2000}
62, 2287; Häkkinen, H.; Yoon, B.; Landman, U.; Li, X.; Zhai, H-J.;
Wang, L-S. \textit{J. Phys. Chem. A} \textbf{2003}, 107, 6168; Remacle,
F.; Kryachko, E. S. \textit{J. Chem. Phys.} \textbf{2005}, 122, 044304;
Ford, M. J.; Hoft, R. C.;  McDonagh, A. \textit{J. Chem. Phys.} \textbf{2005}
109, 20387; de Bas, B. S.; Ford, M. J.; Cortie, M. B. \textit{Theo. Chem.} \textbf{2004} 193, 686;
Wang, J.; Jellinek, J.; Zhao, J.; Chen, Z.; King, B.; von Rague Schleyer, P. \textit{J. Phys. Chem. A} \textbf{2005} 109, 9265. 

\bibitem{cages} Bulusu, S.; Li, X.; Wang, L-S.; Zeng, X. C. \textit{proc. Nat. Acad. Sci.} \textbf{2006} 103, 8326.

\bibitem{science} Li, J.; Li, X.; Zhai, H-J.; Wang, L-S. \textit{Science} \textbf{2003} 299, 864.

\bibitem{Au-melting} Ercolessi, F.; Andreoni, W.; Tosatti, E. \textit{Phys.
Rev. Lett.} \textbf{1991} 66, 911.

\bibitem{Au} Cleveland, C. L.; Luedtke, W. D.; Landman, U. \textit{Phys.
Rev. Lett.} \textbf{1998} 81, 2036.

\bibitem{mike-melt} Soul\'{e} de Bas, B.; Ford, M. J.; Cortie, M. B.
\textit{J. Phys. Condes. Mat.} \textbf{2006} 18, 55.

\bibitem{Ga-prl} Chacko, S.; Joshi, K.; Kanhere, D. G.; Blundell,
S. A. \textit{Phys. Rev. Lett.} \textbf{2004} 92, 135506.

\bibitem{Kavita-PRL} Joshi, K.; Krishnamurty, S; Kanhere, D. G. \textit{Phys.
Rev. Lett.} \textbf{2006} 96, 135703.

\bibitem{Aguado} Aguado, A.; Lopez, J. M. \textit{Phys. Rev. Lett.}
\textbf{2005} 94, 233401.

\bibitem{Our-PRBsn10} Joshi, K.; Kanhere, D. G.; Blundell, S. A.
\textit{Phys. Rev. B} \textbf{2002} 66, 155329.

\bibitem{Our-Silicon} Krishnamurty, S.; Joshi, K.; Kanhere, D. G.;
Blundell, S. A. \textit{Phys. Rev. B.} \textbf{2006} 73, 045419.

\bibitem{James-Tin} Chuang, F.--C.; Wang, C. Z.; Ogut, S.; Chelikowsky,
J. R.; Ho, K. M. \textit{Phys. Rev. B} \textbf{2004} 69, 165408.

\bibitem{Eur-Phys.J} Manninen, K.; Rytkönen, A.; Manninen, M. \textit{Eur.
Phys. J.} \textbf{2004} 29, 39.

\bibitem{Na-PRB} Chacko, S.; Kanhere, D. G.; Blundell, S. A. \textit{Phys.
Rev. B} \textbf{2005} 71, 155407.

\bibitem{Na-JCP} Lee, M. S.; Chacko, S.; Kanhere, D. G. \textit{J.
Chem. Phys.} \textbf{2005} 123, 164310.

\bibitem{Haberland-1997} Schmidt, M.; Kusche, R.; Kronmüller, W.;
von Issendorf, B.; Haberland, H. \textit{Phys. Rev. Lett.} \textbf{1997}
79, 99.

\bibitem{Haberland-Nature} Schmidt, M.; Kusche, R.; von Issendorf,
B.; Haberland, H. \textit{Nature (London)} \textbf{1998} 393, 238.

\bibitem{Haberland-PRL-2003} Schmidt, M.; Donges, J.; Hippler, Th.;
Haberland, H. \textit{Phys. Rev. Lett.} \textbf{2003} 90, 103401.

\bibitem{Haberland-PRL-2005} Haberland, H.; Hippler, T.; Dongres,
J.; Kostko, O.; Schmidt, M.; von Issendorf, B. \textit{Phys. Rev.
Lett.} \textbf{2005} 94, 035701.

\bibitem{Jarold-Tin} Shvartsburg, A.; Jarrold, M. F. \textit{Phys.
Rev. Lett.} \textbf{2000} 85, 2530.

\bibitem{Jarrold-Gallium1} Breaux, G. A.; Benirschke, R. C.; Sugai,
T.; Kinnear, B. S.; Jarrold, M. F. \textit{Phys. Rev. Lett.} \textbf{2003}
91, 215508.

\bibitem{Jarrold-Jacs} Breaux, G. A.; Hillman, D. A.; Neal, C. M.;
Benirschke, R. C.; Jarrold, M. F. \textit{J. Am. Chem. Soc.} \textbf{2004}
126, 8682.

\bibitem{Jarrold-Al} Breaux, G. A.; Neal, C. M.; Cao, B.; Jarrold,
M. F. \textit{Phys. Rev. Lett.} \textbf{2005} 94, 173401.

\bibitem{Jarrold-Tinfrag} Breaux, G. A.; Neal, C. M.; Cao, B.; Jarrold,
M. F. \textit{Phys. Rev. B} \textbf{2005} 71, 073410.

\bibitem{Ga-PRB} S. Krishnamurty, S. Chacko, D. G. Kanhere, 
                   G. A. Breaux, C. M. Neal, and M. F. Jarrold
                    Phys. Rev. B., {\bf 73}, 045406 (2006).

\bibitem{Al-pre} C. M. Neal, A. K. Starace, M. F. Jarrold, K. Joshi, S. Krishnamurty, and D.G. Kanhere,
                      Unpublished results.


\bibitem{KS} Payne, M. C.; Teter, M. P. ; Allen, D. C.; Arias, T.
A.; Joannopoulos, J. D. \textit{Rev. Mod. Phys.} \textbf{1991} 64,
1045.

\bibitem{Nose} Nose, S. \textit{Mol. Phys.} \textbf{1984} 52, 255.

\bibitem{uspp-vanderbilt} Vanderbilt, D. \textit{Phys. Rev. B} \textbf{1990}
41, 7892.

\bibitem{vasp} \textit{Vienna {\em Ab initio } simulation package},
Technische Universität Wien \textbf{1999}; Kresse, G.; Furthmüller,
J. \textit{Phys. Rev. B} \textbf{1996} 54, 11169.

\bibitem{MH} Ferrenberg A. M.; Swendsen, R. H. \textit{Phys. Rev.
Lett.} \textbf{1988} 61, 2635; Labastie, P.; Whetten, R. L. {Phys.
Rev. Lett.} \textbf{1990} 65, 1567.

\bibitem{amv-review} Kanhere, D. G.; Vichare, A.; Blundell, S. A.
{\em Reviews in Modern Quantum Chemistry}, Edited by Sen, K. D.,
World Scientific, Singapore \textbf{2001}. 

\end{thebibliography}
\end{document}